\documentclass[conference]{IEEEtran}
\usepackage{array}
\usepackage{float}
\usepackage{multicol}
\usepackage{multirow}
\usepackage{url}
\usepackage{graphicx}
\usepackage{xcolor}
\usepackage[linesnumbered,ruled,longend]{algorithm2e}
\usepackage{subfigure}
\usepackage{pdfpages}
\usepackage[nolist]{acronym}
\usepackage{multicol}
\usepackage{multirow}
\usepackage{ctable}
\usepackage{amsfonts}
\usepackage{amsmath}
\usepackage{array}
\usepackage{makecell}

\newcolumntype{M}[1]{>{\centering\arraybackslash}m{#1}}

\newboolean{showcomments}
\setboolean{showcomments}{true}
\ifthenelse{\boolean{showcomments}}
{\newcommand{\mynote}[3]{\fbox{\bfseries\sffamily\scriptsize#1}{\textsf{\emph{\color{#3}{#2}}}}}}
{\newcommand{\mynote}[3]{}}

\begin{acronym}
\acro{BF}[BF]{Best Fit}
\acro{WF}[WF]{Worst Fit}
\acro{VM}[VM]{Virtual Machine}
\acro{DC}[DC]{data center}
\acro{QoS}[QoS]{Quality-of-Service}
\acro{VNE}[VNE]{Virtual Network Embedding}
\acro{MILP}[MILP]{Mixed Integer Linear Programming}
\acro{GPU}[GPU]{Graphics Processing Unit}
\acro{AHP}[AHP]{Analytic Hierarchy Process}
\acro{TOPSIS}[TOPSIS]{Technique for Order Preference by Similarity to Ideal Solution}
\acro{PSO}{Particle Swarm Optimization}
\acro{SDN}{Software Defined Networking}
\acro{SIMD}{Single Instruction Multiple Data}
\acro{FCFS}{First Come First Served}
\acro{VN}[VN]{Virtual Network}
\end{acronym}

\begin{document}
\title{Network-Aware Container Scheduling in Multi-Tenant Data Center}
\author{\IEEEauthorblockN{Leonardo R. Rodrigues,$^{\diamond}$
	Marcelo Pasin,$^{\star}$
	Omir C. Alves Jr.,$^{\diamond}$\\
	Charles C. Miers,$^{\diamond}$
	Mauricio A. Pillon,$^{\diamond}$
	Pascal Felber,$^{\star}$
	Guilherme P. Koslovski$^{\diamond}$}
\IEEEauthorblockA{Graduate Program in Applied Computing -- Santa Catarina State University -- Joinville -- Brazil$^{\diamond}$}
\IEEEauthorblockA{University of Neuch\^{a}tel (UniNE) - Institut d'informatique$^{\star}$ -- Switzerland}
}
\maketitle
\begin{abstract}
Network management on multi-tenant container-based \acl{DC} has critical impact on performance.
Tenants encapsulate applications in containers abstracting away details on hosting infrastructures, and entrust \acl{DC} management framework with the provisioning of network \acl{QoS} requirements.
In this paper, we propose a network-aware multi-criteria container scheduler to jointly process containers and network requirements.
We introduce a new \acl{MILP} formulation for network-aware scheduling encompassing both tenants and providers metrics.
We describe two \acs{GPU}-accelerated modules to address the complexity barrier of the problem and efficiently process scheduling requests.
Our experiments show that our scheduling approach accounting for both network and containers outperforms traditional algorithms used by containers orchestrators.   
\end{abstract}
\IEEEpeerreviewmaketitle
\section{Introduction}

Container-based virtualization offers a lightweight mechanism to host and manage large-scale distributed applications for big data processing, edge computing, stream processing, among others. 
Multiple tenants encapsulate applications' environments in containers, abstracting away details of operating systems, library versions, and server configurations.
With containers, \ac{DC} management becomes application-oriented~\cite{Burns:2016} in contrast to server-oriented when using virtual machines.
Several technologies are used to provide connections between containers, such as virtual switches, bridges, and overlay networks~\cite{Suo-2018}.
Yet, containers are a catalyst for network management complexity.
Network segmentation, bandwidth reservation, and latency control are essential requirements to support distributed applications, but container management frameworks still lack appropriate tools to support \ac{QoS} requirements for network provisioning~\cite{Burns:2016}.

We argue that container networking must address at least three communication scenarios, despite the orchestration framework used by the \ac{DC}:
highly-coupled container-to-container communication, group-to-group communication, and containers-to-service communication.
Google's Kubernetes offers a viable solution to group network-intensive or highly coupled containers, by using \emph{pods}.
A pod is a group of one or more containers with shared storage and network, and pods must be provisioned on a single host.
Because the host bus conducts all data transfers within a pod, communication latency is more constant, increasing the network throughput, achieving values superior to default network switching technologies.
However, for large-scale distributed applications, multiple pods must be provisioned and eventually allocated on distinct servers.

This paper advances the field on network-aware container scheduling, a primary management task on container-based \acp{DC}~\cite{Burns:2016}, by jointly allocating compute and communication resources to host network-aware requests.
The network-aware scheduling is analogous to the virtual network embedding (VNE) problem~\cite{Rost:2019}.
Given two graphs, the first one representing user requested containers and all corresponding communications requirements, and the second denoting the \ac{DC} hosting candidates (servers, virtual machines, links, and paths), one must find a map for each vertex and edge from the request graph to a corresponding vertex and edge on \ac{DC} graph.
Vertices and edges carry weights representing process and bandwidth and constraints. 
The combined scheduling of containers and network \ac{QoS} requires a multi-criteria decision, based on conflicting constraints and objectives.

We formally define in this paper the scheduling problem encompassing the network QoS as a \ac{MILP}.
We later propose two \ac{GPU}-accelerated multi-criteria algorithms to process large-scale requests and \ac{DC} topologies.

The paper is organized as follows.
\S\ref{sec:formulation} describes the problem formulation, while \S\ref{sec:milp} defines an optimal \ac{MILP} for joint container and network \ac{QoS} requirements allocation.
Following, \S\ref{sec:milp-eval} presents the evaluation of the proposed \ac{MILP} highlighting the efficiency and limitations of network-aware scheduling.
Then, \S\ref{sec:gpu} describes the implementation of two \ac{GPU}-accelerated algorithms to speed up the scheduling process, and both algorithms are compared with traditional approaches in \S\ref{sec:gpu-evaluation}.
Related work is reviewed in \S\ref{sec:related} and \S\ref{sec:conclusion} concludes. 

\section{Problem Formulation}
\label{sec:formulation}

\subsection{\ac{DC} Resources and Tenants Requests}

Data center resources (bare metal or virtualized) are represented by $G^{s}(N^{s}, E^{s})$, where $N^{s}$ denotes the physical servers, and $E^{s}$ contains all physical links composing the network topology.
A vector is associated with each physical server $u \in N^{s}$, representing the available capacities ($c^{s}_{u}[r]; r \in R$) where $R$ denotes resources as RAM and CPU.
In addition, $bw^{s}_{uv}$ represents the available bandwidth between physical servers $u$ and $v$.
Thus, a tenant request is given by $Req(N^{c}, E^{c})$, with $N^{c}$ being a set of containers and $E^{c}$ the communication requirements among them.
Also, as in Kubernetes, each container is associated with a pod.

Containers from a pod must be hosted by the same physical server (sharing the IP address and port space).
A group of pods $G$ is defined in a tenant's  request, and a container $i \in N^{c}$ is connected to a pod group $g \in G$, indicated by $i \in pod_{g}$.
Instead of requesting a fixed configuration for each \ac{QoS} requirement, containers are specified as minimum and maximum intervals.
For a container $i$, the minimum and maximum values for any $r \in R$ are respectively defined as $c^{min}_{i}[r]$ and $c^{max}_{i}[r]$.
The same rationale is applied to containers interconnections ($E^{c}$): minimum and maximum bandwidth requirements are given by $bw^{min}_{ij}$ and $bw^{max}_{ij}$.

A container orchestration framework has to determine whether to accept or not a tenant request.
The allocation of containers onto a \ac{DC} is decomposed into nodes and links assignments.
The mapping of containers onto nodes is given by $\mathcal{M}_{c}: N^{c} \mapsto N^{s}$, while the mapping of networking links between containers onto paths is represented as $\mathcal{M}_{ec}: E^{c} \mapsto P^{s}$.
Table~\ref{tab:notation} summarizes the notation used is this paper.

\begin{table}[t!]
\renewcommand{\arraystretch}{0.7}
\scriptsize
    \centering
    \begin{tabular}{m{2cm} m{5.7cm} }
        \textbf{Notation}                 & \textbf{Description} \\
        \toprule
        $G^{s}(N^{s}, E^{s})$             & \ac{DC} graph composed of $N^{s}$ servers and $E^{s}$ links.\\
        \midrule
        $c^{s}_{u}[r]$                      & Resource capacity vector of server $u \in N^{s}$.\\
        \midrule
        $P^{s}$                          & All direct paths (physical and logical) on \ac{DC} topology.\\
        \midrule
        $bw^{s}_{uv}$                      & Bandwidth capacity between servers $u$ and $v$, $uv \in E^{s}$. \\
        \midrule
        $Req (N^{c}, E^{c})$             & Request, composed of $N^{c}$ containers and $E^{c}$ links.\\
        \midrule
        $c^{min}_{i}[r]$,  $c^{max}_{i}[r]$            & Minimum and maximum resources capacities for container $i \in N^{c}$.\\
        \midrule     
        $bw^{min}_{ij}$, $bw^{max}_{ij}$               & Minimum and maximum bandwidth requirement between containers $i$ and $j$, $ij \in E^{c}$.\\
        \midrule
        $pod_{g} \subset N^{c} $              & Set of containers $i \in N^{c}$ composing a pod $g \in G$. \\
        \bottomrule
    \end{tabular}
    \caption{Notation used along this paper: $i$ and $j$ are used for indexing containers, while $u$ and $v$ are used for \ac{DC} servers.}
    \label{tab:notation}
\end{table}

\subsection{Objectives}
\label{sec:objectives}

\noindent\textbf{Energy consumption.}
To reduce energy consumption, we pack containers in as few nodes as possible, allowing to power off the unused ones.
We call this technique \emph{consolidation}, and we reach it by minimizing the \ac{DC} fragmentation, defined as the ratio of the number of active servers (\textit{e.g.}, those hosting containers) to the total number of \ac{DC} resources.
Server fragmentation is given by $\mathcal{F}(N^{s}) = |N^{s'}| / |N^{s}|$, while the same rationale is applied for links, $\mathcal{F}(E^{s}) = |E^{s'}| / |E^{s}|$, where $|N^{s'}| $ and $|E^{s'}|$ denote the number of active servers and links, respectively.

\noindent\textbf{\acl{QoS}.}
A container can be successfully executed with any capacities configuration in the intervals specified as minimum and maximum.
However, optimal performance is reached when the maximum values are used.
In this sense, utility functions can be applied for representing the improvement on container's configuration.
In short, the goal is to maximize Eq.~(\ref{eq:objective-utility-container}) and~(\ref{eq:objective-utility-container-link}) for each container $i \in N^{c}$, where $c^{a}_{iu}[r]$ and $bw^{a}_{ijuv}$ represent the capacity effectively allocated for vertices and edges, respectively.

\small
\begin{eqnarray}
    \label{eq:objective-utility-container}
    \mathcal{U}(i) = \frac{\sum_{r \in R} \frac{c^{a}_{iu}[r]}{c^{max}_{i}[r]}}{|R|} ; u = \mathcal{M}_{c}(i) \\
    \label{eq:objective-utility-container-link}
    \mathcal{U}(ij) = \frac{\sum_{uv \in \mathcal{M}_{e}(ij)} bw^{a}_{ijuv}}{bw^{max}_{ij}} 
\end{eqnarray}
\normalsize

\section{Optimal MILP for Joint Container and Network \ac{QoS} Allocation}
\label{sec:milp}

\subsection{Variables and Objective Function}
\label{sec:objective_function}
A set of variables (Table~\ref{tab:variables}) are proposed to find a solution for joint allocation of containers and bandwidth requirements, as well as to achieve maps $\mathcal{M}_{c}: N^{c} \mapsto N^{s}$, and $\mathcal{M}_{ec}: E^{c} \mapsto P^{s}$.
The binary variable $x_{iu}$ accounts the mapping of containers on servers.
The containers' connectivity ($xl_{ijuv}$) applies the same rationale.
For identifying the amount of resources allocated to a container $i \in N^{c}$, the float vector $c^{a}_{i}$ is introduced.
Bandwidth allocation follows the same principle and is accounted by float variable $bw^{a}_{ij}$.

\begin{table}[htb!]
\renewcommand{\arraystretch}{0.7}
  \centering
  \begin{tabular}{llm{57mm} }
  \textbf{Notation}& \textbf{Type} & \textbf{Description} \\\toprule
  $x_{iu}$         & Bool & Container $i \in N^{c}$ is mapped on server $u \in N^{s}$.\\\midrule
  $xl_{ijuv}$      & Bool & Connection $ij \in E^{c} $ is mapped on link $uv \in E^{S}$.\\\midrule
  $c^{a}_{iu}[r]$  & Float   & Resource ($r \in R$) capacity vector allocated to container $i \in N^{c}$ on server $u \in N^{s}$.\\\midrule
  $bw^{a}_{ijuv} $ & Float   & Bandwidth allocated to connection $ij \in E^{c}$ on link $uv \in E^{s}$.\\\midrule
  $f_{u}$          & Bool & Server $u \in N^{s}$ is hosting at least one container.\\\midrule
  $fl_{uv} $       & Bool & Link $uv \in E^{s}$ is hosting at least one connection.\\\bottomrule
    \end{tabular}
    \caption{MILP variables for mapping containers and virtual links atop a multi-tenant \ac{DC}.}
    \label{tab:variables}
\end{table}

The objectives (\S\ref{sec:objectives}) are reached by the minimization of Eq.~(\ref{eq:milp-objective}).
Two additional binary variables are used to identify if \ac{DC} resources are hosting at least one container or link, $f_{u}$ and $fl_{uv}$.
Value $1$ is set just for active servers, as given by $f_{u} \geq \frac{\sum_{i \in N^{c}} x_{iu}}{|N^{c}|} ; \forall u \in N^{s}$.
Physical links follow the same idea, $fl_{uv} \geq \frac{\sum_{ij \in E^{c}} xl_{ijuv}}{|E^{c}|} ; \forall uv \in E^{s}$.
Finally, the importance level of each term is defined by setting $\alpha$.

\scriptsize
\begin{eqnarray}
    \label{eq:milp-objective}
    minimize: 
       \alpha \bigg( \sum_{i \in N^{c}} (1 - \mathcal{U}(i)) + \sum_{ij \in E^{c}} (1 - \mathcal{U}(ij))  \bigg) \nonumber \\
    + (1 - \alpha) \bigg( \sum_{u \in N^{s}} \frac{f_{u}}{|N^{s}|} + \sum_{uv \in E^{s}} \frac{fl_{uv}}{|E^{s}|} \bigg)
\end{eqnarray}
\normalsize

\subsection{Constraints}
\label{sec:constraints}

\noindent\textbf{\ac{DC} Capacity, \ac{QoS} Constraints and Integrity of Pods.}
A \ac{DC} server $u \in N^{s}$ must support all hosted containers, as indicated by Eq.~(\ref{eq:constraint-vm-cpu}), while the bandwidth of link $uv \in E^{s}$ must support all containers transfers allocated to it, as given by Eq.~(\ref{eq:constraint-vm-links}).
Eq.~(\ref{eq:constraint-container-cpu}) guarantees the allocation of a resources capacities from min-max intervals for a containers $i \in N^{c}$.
The same rationale is applied for $ij \in E^{c}$ on Eq.~(\ref{eq:constraint-containers-links}).

\small
\begin{eqnarray}
    \label{eq:constraint-vm-cpu}
    c^{s}_{u}[r] \geq \sum_{i \in N^{c}} c^{a}_{iu}[r] ; \forall u \in N^{s} ; \forall r \in R \\
    \label{eq:constraint-vm-links}
    bw^{s}_{uv} \geq \sum_{ij \in E^{c}} bw^{a}_{ijuv} ; \forall uv \in E^{s} \\
    \label{eq:constraint-container-cpu}
    c^{min}_{i}[r] \times x_{iu} \leq c^{a}_{iu}[r] \leq c^{max}_{i}[r] \times x_{iu} \nonumber \\ \forall i \in N^{c}; \forall u \in N^{s} ; \forall r \in R \\
    \label{eq:constraint-containers-links}
    bw^{min}_{ij} \times xl_{ijuv} \leq bw^{a}_{ijuv} \leq bw^{max}_{ij} \times xl_{ijuv} \nonumber \\ \forall ij \in E^{c}; uv \in E^{s} \\
    \label{eq:constraint-container-pod}
    x_{iu} = x_{ju}        ; \forall g \in G ; \forall i, j \in pod_{g} ; \forall u \in N^{s}   
\end{eqnarray}
\normalsize

Finally, containers are optionally organized in pods.
For guaranteeing the integrity of pods specifications, Eq.~(\ref{eq:constraint-container-pod}) indicates that all resources from a pod ($i, j \in pod_{g}$) must be hosted by the same server ($u \in N^{s}$).

\noindent\textbf{Binary and Allocation Constraints.}
A container must be hosted by a single server ($\sum_{u \in N^{s}} x_{iu} = 1 ; \forall i \in N^{c}$), while each virtual connectivity between containers is mapped to a path between resources hosting its source and destination as given by $\sum_{v \in N^{s}} xl_{ijvu} + \sum_{v \in N^{s}} xl_{ijuv} = x_{iu} + x_{ju};  \forall u \in N^{s}; \forall ij \in E^{c}$.
However, on large scale \ac{DC} topologies, servers are interconnected by multiple paths composed of at least one switch hop.
In order to keep the model realistic with current \acp{DC}, we rely on network management techniques, such as \acs{SDN}~\cite{Souza-2017} to control the physical links usage and populate the $E^{s}$ with updated information and available paths.

\section{Evaluation of the Optimal MILP for Network-Aware Containers Scheduling}
\label{sec:milp-eval}

The \ac{MILP} scheduler and a discrete event simulator were implemented in Python $2.7.10$ using CPLEX optimizer ($v12.6.1.0$).
For composing the baseline was used the native algorithms offered by containers orchestrators, \ac{BF} (binpacking) and \ac{WF} (spread).
As \ac{BF} and \ac{WF} natively ignore the network requirements, we included a shortest-path search after the allocation of servers to host containers for conducting a fair comparison.

\subsection{Metrics and \ac{MILP} Parametrization}

The \ac{MILP} objective function, Eq.~(\ref{eq:milp-objective}), is composed of terms to represent the tenant's perspective (the utility of network allocation and the queue waiting time) and the \ac{DC} fragmentation (the provider's perspective).
Although a minimum value is requested for each container parameter, the optimal utility function expects the allocation of maximum values ($\mathcal{U}(.) = 1$).
The \ac{MILP}-based scheduler is guided by the $\alpha$ value to define the importance of each term composing the objective function.
For demonstrating the impact of defining $\alpha$, we evaluated $3$ configurations $\alpha = 0; 0.5; 1$.
Configurations with $\alpha = 0$ and $\alpha = 1$ define the baseline for comparisons: by setting $\alpha = 0$ the \ac{MILP} optimizes the problem regarding the fragmentation perspective only, while $\alpha = 1$ represents the opposite; more importance is given to containers and network utilities. 

\subsection{Experimental Scenarios}

\subsubsection{DC Configuration}

A Clos-based topology (termed Fat-Tree) is used to represent the \ac{DC}~\cite{Jupiter-2015, NiranjanMysore:2009}.
The $k$ factor guides the topology indicating the number of switches, links, and hosts used to compose the \ac{DC}.
A fat-tree build with $k$-port switches supports up to $k^{3}/4$ servers.
The \ac{DC} is configured with $k=4$, and composed of homogeneous servers equipped with $24$ cores and $256$ GB RAM, while the bandwidth capacity for all links is defined as $1$ Gbps.

\subsubsection{Requests}

A total of $200$ requests is submitted with resources specifications based on uniform distributions for containers capacities, submission time, and duration.
Each request is composed of $5$ containers with a running time up to $200$ events from a complete execution of $500$ events.
For composing the pods, up to $50\%$ of containers from a single requested are grouped in pods.
For the network, the bandwidth requirement between a pair of containers is configured up to $50$ Mbps, besides requests with $1$ Mbps requirement representing applications without burdensome network requirements.
The values for CPU and RAM configuration are uniformly distributed up to $2$ and $4$, respectively. 

\subsection{Results and Discussion}

\begin{figure}
	\begin{center}
	\subfigure[\ac{DC} network fragmentation.]{
            \includegraphics[width=\linewidth]{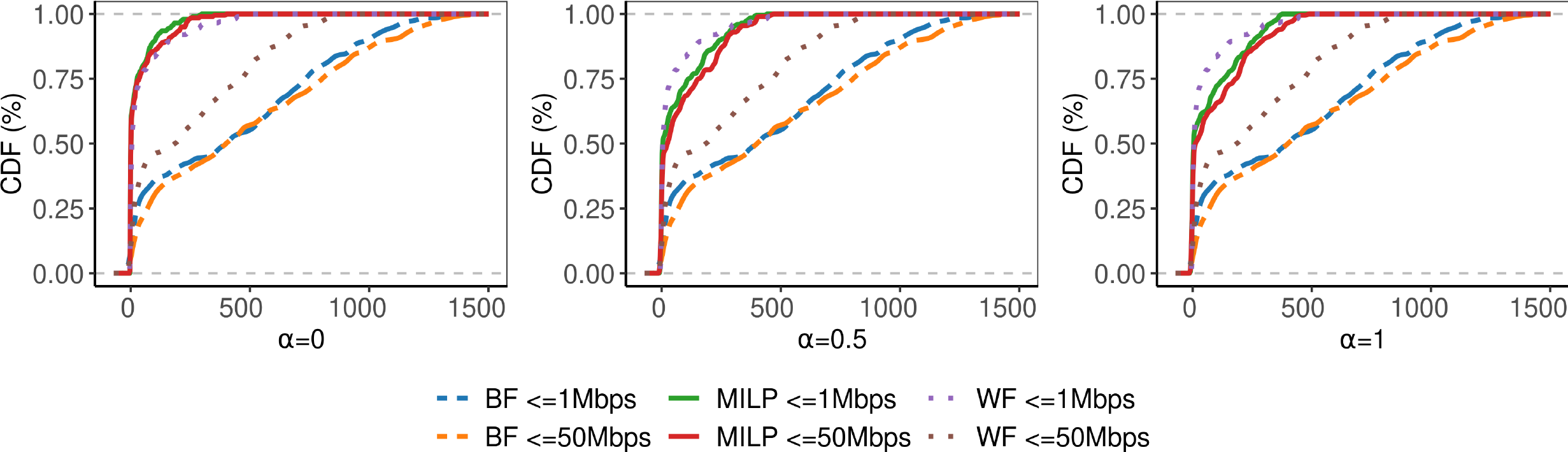}
			\label{fig:delay-pod50}
	}
	\subfigure[\ac{DC} links fragmentation.]{
            \includegraphics[width=\linewidth]{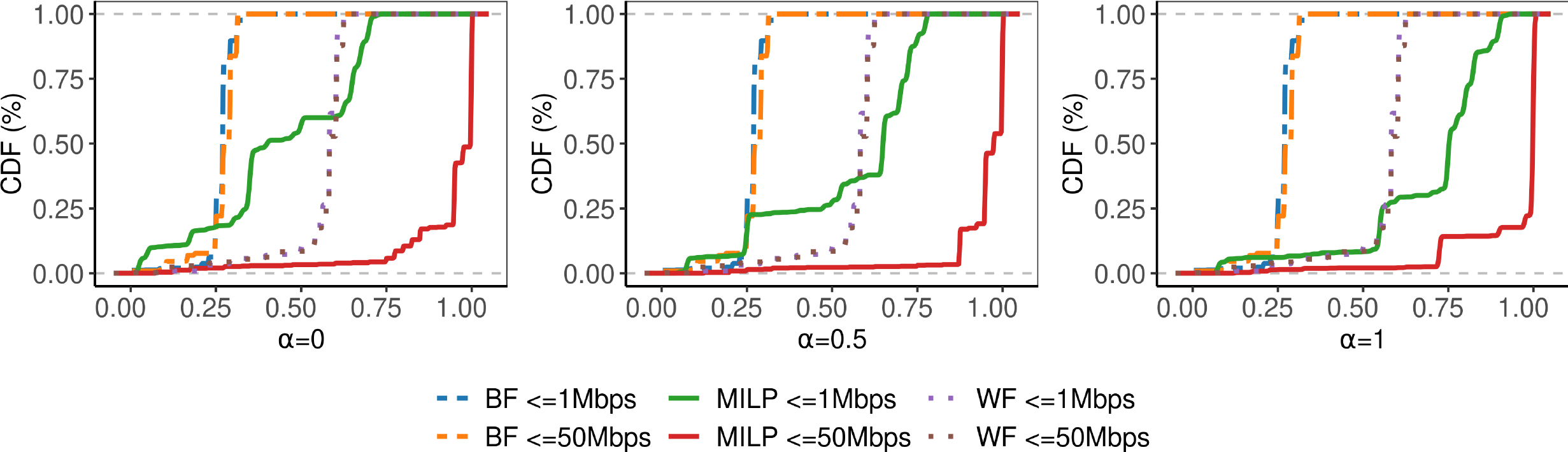}
            \label{fig:lf-pod50}
     }
     \end{center}    
     \vspace{-0.2cm}
     \caption{Request utility and delay, and \ac{DC} fragmentation when executing the \ac{MILP}-based scheduler.}
\end{figure}

Table~\ref{tab:utility-milp} and Figures~\ref{fig:delay-pod50} and~\ref{fig:lf-pod50} present results for utility of network and container requests, provisioning delays, and \ac{DC} network fragmentation, respectively.

\begin{table}[b]
\scriptsize
    \centering
    \begin{tabular}{ c c r r r }
        \textbf{\thead{Algorithm}} & \textbf{$\alpha$} &\textbf{\thead{Bandwidth}} & \textbf{\thead{$\mathcal{U}(ij)$ }}  &   \textbf{\thead{$\mathcal{U}(i)$}}\\
    \toprule
    \multirow{6}{*}{MILP} & \multirow{2}{*}{0} & 1 Mbps & $22.68\%$& $99.90\%$\\ \cline{3-5}
    & & 50 Mbps & $7.86\%$& $66.29\%$\\ \cline{2-5}
    
    & \multirow{2}{*}{0.5} & 1 Mbps & $26.78\%$ & $99.90\%$\\ \cline{3-5}
    & & 50 Mbps & $86.67\%$ & $97.21\%$\\ \cline{2-5}
    
    & \multirow{2}{*}{1} & 1 Mbps & $38.28\%$ & $99.90\%$ \\ \cline{3-5}
    & & 50 Mbps & $93.56\%$& $97.80\%$\\ 
    \midrule    
    
    \multirow{2}{*}{WF} & \multirow{2}{*}{-} & 1 Mbps & $100\%$ & $98.03\%$ \\ 
    \cline{3-5}
    &  & 50 Mbps & $100\%$ & $99.98\%$ \\
    \midrule    

    \multirow{2}{*}{BF} & \multirow{2}{*}{-} & 1 Mbps &  $100\%$ & $97.20\%$\\ \cline{3-5}
     & & 50 Mbps & $100\%$ & $99.46\%$\\
    \bottomrule
    \end{tabular}
    \vspace{-0.2cm}
    \caption{Link and container utilities for \ac{MILP}, \ac{BF}, and \ac{WF}.}
    \label{tab:utility-milp}
\end{table}

\vspace{-0.2cm}
\ac{BF} and \ac{WF} algorithms have a well-defined pattern for all network utility metric.
For requests with low network requirements (up to $1$Mbps), both algorithms tend to allocate the maximum requested value for network \ac{QoS}.
An exception is observed for \ac{BF} with network-intensive requests (up to $50$Mbps) as the algorithm gives priority to minimum requested values for consolidating requests on \ac{DC} resources.
With regarding the network-aware \ac{MILP} scheduler, even for requests with $\alpha = 0$ focusing on decreasing the \ac{DC} fragmentation, the scheduler allocated maximum values for network requests, following the \ac{BF} and \ac{WF} algorithms.
However, the impact of $\alpha$ parametrization is perceived for network-intensive requests.
The \ac{MILP} configuration with $\alpha = 0.5$ shows that the algorithm can jointly consider requests utility and \ac{DC} fragmentation. 
The results in Fig.~\ref{fig:lf-pod50} show that scheduling network-intensive requests increases the network \ac{DC} fragmentation.
The provisioning delays~(Figure~\ref{fig:delay-pod50}) explain this fact: the \ac{MILP} scheduler decreases the queue waiting time for network-intensive requests when compared to \ac{BF} and \ac{WF}.

In summary, it is evident that network \ac{QoS} must be considered by the scheduler to decrease the queue waiting time and to reserve utility's dynamic configurations.
Moreover, the results obtained from MILP configured with $\alpha = 0.5$ demonstrated the real trade-off between fragmentation and utility, or in other words, provider's and tenant's perspectives. 

\section{GPU-Accelerated Heuristics}
\label{sec:gpu}

Although \ac{MILP} is efficient to model and highlights the impact of network-aware scheduling, solving this problem is known to be computationally intractable~\cite{Rost:2019} and practically infeasible for large-scale scenarios.
Therefore, we developed two \ac{GPU}-accelerated multi-criteria algorithms to speed up the joint scheduling of containers and network with \ac{QoS} requirements.
We selected two multi-criteria algorithms: \ac{AHP} and \ac{TOPSIS}, chosen due to their multidimensional analysis, being able to work with several servers simultaneously.
Also, \ac{AHP} and \ac{TOPSIS} provide a structured method to decompose the problem and to consider trade-offs in conflicting criteria.
Following the notation used to express the \ac{MILP} (Table~\ref{tab:notation}), both algorithms analyze the same set of criteria $c^{s}_{u}$ for a given server $u$.
In addition, the sum of all bandwidth capacity $bw^{s}_{uv}$ with source on $u$ (given by $bw^{s}_{u}$) and the current server fragmentation ($f_{u}$) are accounted and included on $c^{s}_{u}$ capacity vector. 
The multi-criteria algorithms analyzed all variables described in Section~\ref{sec:objectives} as attributes.

\subsection{Weights Distribution}

\ac{AHP} and \ac{TOPSIS} algorithms are guided by a weighting vector to define the importance of each criteria. 
While the \ac{MILP} has $\alpha$ to indicate the importance level of each term in the objective function, the multi-criteria function decomposes $\alpha$ into a vector $W = \{\alpha_{0},\alpha_{1},...\alpha_{|R| - 1}\};  \sum_{i \in R} \alpha_{i} = 1$. 
Tab.~\ref{tab:multi_weights} presents different $W$ compositions to the \ac{MILP} objective.

\begin{table}[htbp!]
\renewcommand{\arraystretch}{0.9}
\scriptsize
    \centering
    \begin{tabular}{M{1.2cm} M{1.cm} M{1cm} M{1.5cm} M{1.cm}}
        \textbf{Scenario}  &  \textbf{CPU} & \textbf{RAM} & \textbf{Fragmentation} & \textbf{Bandwidth}\\
        \toprule
        Flat                 & 0.25 & 0.25 & 0.25   & 0.25\\
        \midrule
        Clustering    & 0.17   & 0.17&  0.5  & 0.16   \\
        \midrule
        Network        &0.17    & 0.17&  0.16    & 0.5 \\
        \bottomrule
    \end{tabular}
    \caption{Weighting schema for \ac{AHP} and \ac{TOPSIS}. The Flat configuration is equivalent to $\alpha = 0.5$ in \ac{MILP}, while Clustering and Network represents $\alpha = 0$ and $\alpha = 1$, respectively.}
    \label{tab:multi_weights}
\end{table}

The multi-criteria analysis with clustering configuration optimizes the problem aiming at \ac{DC} consolidation (equivalent to $\alpha = 0$  on \ac{MILP} formulation) through the definition of high importance level ($50\%$) to fragmentation criteria, while the other criteria share equally the last $50\%$. 
In other hand, the execution with network configuration ($\alpha = 1$ from \ac{MILP} formulation), the bandwidth criteria receive a higher importance level ($50\%$) while the other criteria share equally the last $50\%$. 
This configuration makes the scheduler select servers that have the highest residual bandwidth.
Finally, the flat configuration sets the same importance weight for all criteria (following the $\alpha = 0.5$ rationale on \ac{MILP}).

\subsection{\ac{AHP}}

The \ac{AHP} is a multi-criteria algorithm that hierarchically decomposes the problem to reduce the complexity, and performs a pairwise comparison to rank all alternatives~\cite{saaty2005making}.
In short, the hierarchical organization is composed of three main levels.
The objective of the problem is configured at the top of the hierarchy, while the set of criteria is placed in the second level, and finally, in the third level represents all the viable alternatives to solve the problem. 

In our context, the selection of the most suitable \ac{DC} to host a container is performed in steps.
In the first step two vectors ($M_{1}$ and $M_{2}$) are built combining all criteria and alternatives (second and third level of \ac{AHP} hierarchy) applying the weights defined in Table~\ref{tab:multi_weights}.
In other words, $M_{1}[v] =  W[v] ; \forall v \in R$ while $M_{2}[v \times |N^{s}| + u] =  c^{s}_{u}[v] ; \forall u \in N^{s} ; \forall v \in R$.
The representation based on a vector was chosen to exploit the \ac{SIMD} \ac{GPU}-parallelism.
Later the pairwise comparison is applied for all elements into the hierarchy.
If $M_{1}[v \times |R| + u] > 0$, the value $M_{1}[v \times |R| + u] - M_{1}[i \times |R| + u]$ is attributed to;
In addition, if the cell value is $<0$, $\frac{1}{M_{1}[v \times |R| + u] - M_{1}[i \times |R| + u]}$ is set; and $1$ otherwise.
The same rationale is applied for $M_{2}$, indexed by $v \times |N^{s}|^{2} + i \times |N^{s}| + u$.
Later, both vectors are normalized.
At this point, the algorithm calculates the local rank of each element in the hierarchy ($L_{1}$ and $L_{2}$), as described in Eqs.~(\ref{eq:pmlAHP_L1}) and (\ref{eq:pmlAHP_L2}), $\forall u, v \in R; \forall i, j \in N^{s}$.
Finally, the global priority ($PG$) of the alternatives is accounted to guide the host selection, as given by $PG[v] = \sum_{x \in N^{s}}P_{1}[v] \times P_{2}[v \times |N^{s}| + x]$.

\small
\begin{eqnarray}
    \label{eq:pmlAHP_L1}
    L_{1} [v \times |R| + u] = \frac{\sum_{x \in R} M_{1}[v \times |R| + x]}{|R|}  \\
    \label{eq:pmlAHP_L2}
    L_{2}[v \times |N^{s}| + j] = \frac{\sum_{x \in N^{s}} M_{2}[v \times |N^{s}|^{2} + i \times |N^{s}| + x]}{|N^{s}|}    
\end{eqnarray}
\normalsize

\subsection{\ac{TOPSIS}}

The \acf{TOPSIS} is based in the shortest Euclidean Distance from the alternative to the ideal solution~\cite{Hwang-1981}.
The benefits of this algorithm are three-fold: \textit{(i)} can handle a large number of criteria and alternatives; \textit{(ii)} requires a small number of qualitative inputs when compared to \ac{AHP}; and \textit{(iii)} is a compensatory method, allowing the analysis of trade-off criteria.

The ranking of the \ac{DC} candidates is performed in steps.
Initially, the evaluation vector $M$ correlates \ac{DC} resources ($N^{s}$) and the criteria elements ($R$): $M[v \times |N^{s}| + u] =  c^{s}_{u}[v] ; \forall u \in N^{s} ; \forall v \in R$, which is later normalized.
The next step is the application of weighting schema on $M$ values: $M[v \times |N^{s}| + u] = M[v \times |N^{s}| + u]  \times W[v]  ;\forall u \in N^{s};\forall v \in R$.
Based on $M$, two vectors are them composed with the maximum and minimum values for each criteria, represented by $A^{+}$ (the upper-bound solution quality) and $A^{-}$ (the lower-bound).
\ac{TOPSIS} requires the calculation of Euclidean distances between $M$ and upper- and lower-bounds, composing $Ed^{+}$ and $Ed^{-}$.
Finally, a closeness coefficient array is accounted for all \ac{DC} servers, $Rank[u] = \frac{Ed^{-}[u]}{Ed^{+}[u]+Ed^{-}[u]}; \forall u =  N^{s}$, and afterwards the resulting array is sorted on decreasing order, indicating the selected candidates. 

\subsection{\ac{GPU} Implementation}

The \ac{AHP} and \ac{TOPSIS} are decomposed in \ac{GPU}-tailored kernels following a pipeline execution.
The first kernel is in charge is acquiring \ac{DC} and network-aware containers requests, while the remaining kernels perform the comparisons using the parallel reduction technique.
A special explanation is required for selecting physical paths to host containers interconnections.
After the selection of the most suitable server for each pod presented in the tenant's request, the virtual links between the containers must be set. 
A modified Dijkstra algorithm is used to compute the shortest path that has the maximum available bandwidth between the hosting servers.
The modified Dijkstra is implemented as a single kernel to allow multiple executions, where each thread calculates a different source and destination pair.
As the links between every two nodes in the \ac{DC} are undirected, the \ac{GPU} implementation uses a specific array representation to reduce the total space needed.
The main principle of the data structure of this algorithm is that $u < v$ where $u$ is the source and $v$ the destination, and the paths $u \rightarrow v$ and $v \rightarrow u$ are the same.

\section{Evaluation of GPU-Accelerated Heuristics}
\label{sec:gpu-evaluation}

The GPU-accelerated scheduler and a discrete event simulator were implemented in C++, using GCC compiler $v. 8.2.1$ and CUDA framework $v. 10.1$.

\subsection{Experimental Scenarios}
\label{sub:experimental}

The evaluation considers a \ac{DC} composed of of homogeneous servers equipped with $24$ cores, $256$ GB RAM and interconnected by a Fat-Tree topology ($k=20$) and bandwidth capacity of $1$ Gbps for all links.
A total of $6000$ requests were submitted to be scheduled, each composed of $4$ containers with a running time up to $250$ events from a complete execution of $500$ events.
For composing the requests, up to $50\%$ of containers from a single request are grouped in pods, while the bandwidth requirement between a pair of containers is configured up to $50$ Mbps (a heavy network requirement).

\subsection{Results and Discussion}

\begin{figure}[htb!]
	\begin{center}    
    \subfigure[\ac{DC} network fragmentation from the GPU-accelerated scheduler.]{
            \includegraphics[width=\linewidth]{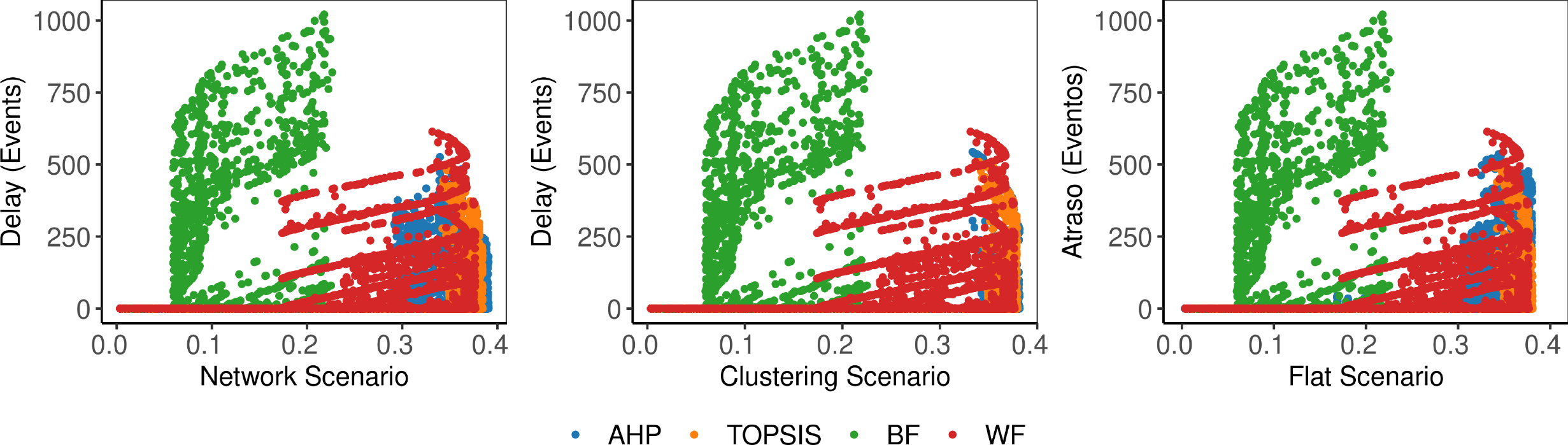}
			\label{fig:gpu-delayxlinkload}
    }
     \subfigure[\ac{DC} links fragmentation from the GPU-accelerated scheduler.]{
            \includegraphics[width=\linewidth]{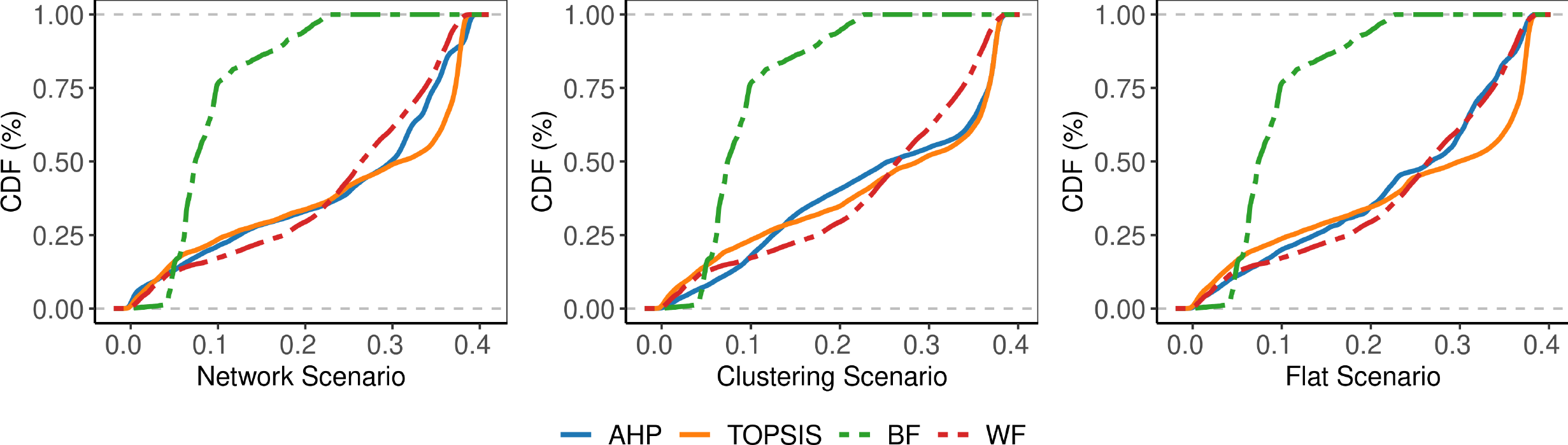}
			\label{fig:gpu-linkfrag}
    }
    \end{center}
     \caption{Requests utility and delay, and \ac{DC} fragmentation when executing the \ac{GPU}-based scheduler.}
\end{figure}

Results are summarized by Table~\ref{tab:summary_gpu} and Figures~\ref{fig:gpu-delayxlinkload} and~\ref{fig:gpu-linkfrag}, showing data for the runtime, utility of network and container requests, provisioning delays correlated to the \ac{DC} fragmentation and \ac{DC} network fragmentation, respectively.

\begin{table}[htb!]
\scriptsize
\renewcommand{\arraystretch}{1.1}
    \centering
	\resizebox{\columnwidth}{!}{%
    \begin{tabular}{c  l r r r r}
       \textbf{Algorithm} & \textbf{\thead{Scenario}}  &   \textbf{\thead{$\#$ Events}} & \textbf{\thead{Average Runtime (s)}} & \textbf{\thead{$\mathcal{U}(ij)$}} & \textbf{$\mathcal{U}(i)$}\\
	\toprule
	BF & - & $2462$ & $79.38$ &$100\%$ & $96.89\%$ \\
         \midrule
         WF  & - & $1007$ & $47.80$ & $100\%$&$99.41\%$\\
         \midrule
         \multirow{3}{*}{AHP} & Flat & $949$ & $9.45$ &$100\%$&$98.22\%$\\
         \cline{2-6}          
        & Clustering & $936$ & $7.51$  &$100\%$&$99.10\%$\\
        \cline{2-6}         
        & Network &  $928$ & $6.90$  &$100\%$&$98.41\%$\\
         \midrule
        \multirow{3}{*}{Topsis} & Flat & $894$ & $3.67$ &$100\%$&$98.85\%$ \\
         \cline{2-6}
        & Clustering & $916$ & $3.84$  &$100\%$&$99.01\%$\\
         \cline{2-6}          
        & Network & $892$ & $3.48$  &$100\%$&$98.94\%$\\
        \bottomrule
            
    \end{tabular}
    }
    \caption{Runtime, Link and Container Utilities for \ac{BF}, \ac{WF}, AHP and TOPSIS.}
    \label{tab:summary_gpu}
\end{table}

\vspace{-1.1cm}

\vspace{-0.3cm}
Figure~\ref{fig:gpu-delayxlinkload} shows that the multi-criteria algorithms have a small variation for request delay, grouping the data in high fragmentation percentages, while the \ac{WF} induces delay in requests regardless the \ac{DC} fragmentation.
In turn, the \ac{BF} algorithm imposes higher delay to requests resulting in a small fragmentation percentage, below $30\%$ of network fragmentation. 
\ac{WF} and \ac{BF} generate a long requests queue impacting directly in the total computational time needed to schedule all the tenants' requests.

Regarding the container's utility (Table~\ref{tab:summary_gpu}), the multi-criteria algorithms give priority to schedule requests mixing between the maximum and minimum requirements, increasing the number of containers in the \ac{DC}. 
The \ac{WF} tends to allocate the maximum value for the requests, while the \ac{BF} tends to give the minimum values of the requests.
While the multi-criteria algorithms increase the number of containers in the \ac{DC} reducing the total delay, the network fragmentation have similar behavior with the \ac{WF} algorithm, as shown in Figure~\ref{fig:gpu-linkfrag}. 
Meanwhile the \ac{BF} keeps the network fragmentation small due to the long delays that it applies in the requests.
It is possible to observe that the multi-criteria algorithms present better consolidation results when compared to the \ac{WF} and \ac{BF} algorithms, due to their capacity to allocate more requests in the \ac{DC} keeping the fragmentation similar to \ac{WF}. 
It is possible to conclude that the network weighting schema is essential to perform a joint scheduling of container and network requirements.
It is important to emphasize: the \ac{GPU}-accelerated algorithms can schedule the requests with bandwidth requirements atop a large-scale \ac{DC} in a few seconds. 
Specifically, \ac{TOPSIS} outperformed \ac{BF}, \ac{WF}, and \ac{AHP} results.

\section{Related Work}
\label{sec:related}

The orchestration and scheduling of virtualized \ac{DC} is a trendy topic of the specialized literature.
\ac{MILP} techniques offer optimal solutions which are generally used as a baseline for comparisons~\cite{Souza-2017}, but the problem complexity and search space often create opportunities for heuristic-based solutions~\cite{Rost:2019}.

Guerrero \cite{Guerrero-2018} proposed a scheduler for container-based micro-services.
The containers workload and the networking details were analyzed to perform the \ac{DC} load balance.
Guo \cite{guo2018container} proposed a scheduler to optimize the load balancing and workload through the neighborhood division in a micro-service method.
Both proposals were analyzed on small-scale \acp{DC} as the problem complexity imposes a barrier on real-scale use.
The GenPack~\cite{Havet-2017} scheduler employs monitoring information to define the appropriated group of a container based on the resource usage, avoiding resources disputes among containers.
A security-concerned scheduler was proposed by \cite{Vaucher-2018}, based on bin-packing executing a \ac{BF} approach.
\ac{GPU}-accelerated algorithms can be applied to speed-up these heuristics reaching large-scale \acp{DC}~\cite{NesiCNSM:2018}.

A joint scheduler based on priority-queue, \ac{AHP} and \ac{PSO} is proposed by~\cite{alla2016efficient}.
The requests are sorted by their priority level and waiting time, and then the tasks are sent to the \ac{AHP} to be ranked and then serving as an input to \ac{PSO}.
The results show a reduction on makespan up to $15\%$ when compared to \ac{PSO}.
In addition, \cite{panwartopsis} proposed a \acs{VM} scheduler based on \ac{TOPSIS} and \ac{PSO}.
The scheduler was compared with $5$ meta-heuristics using the $4$ metrics: makespan, transmission time, cost and resource utilization, achieving an improvement up to 75\% when compared to traditional schedulers.
Although many multi-criteria solutions appear in the literature, we were unable to find schedulers dealing with containers, pods, and their virtual networks.

Network requirements are disregarded or partially attended by major of reviewed schedulers.
Even well-known orchestrators (\textit{e.g.}, Kubernetes) consider the network as second-level and not critical parameters.
Containers are used to model large-scale distributed applications, and it is evident that network allocation can impact on applications performance~\cite{Suo-2018}.
\section{Conclusion}
\label{sec:conclusion}

We investigated the joint scheduling of network \ac{QoS} and containers on multi-tenant \acp{DC}.
A \ac{MILP} formulation and experimental analysis reveal that a network-aware scheduler can decrease \ac{DC} network fragmentation and processing delays.
However, solving a \ac{MILP} is known to be computationally intractable and practically infeasible for large-scale scenarios.
We then developed two \ac{GPU}-accelerated multi-criteria algorithms, \ac{AHP} and \ac{TOPSIS}, to schedule requests on a large-scale \ac{DC}.
Both network-aware algorithms outperformed the traditional schedulers with regard to \ac{DC} and tenant perspectives.
Future work includes the scheduling of batch requests and a distributed implementation for increasing the fault tolerance.

\section*{Acknowledgments}
The research leading to the results presented in this paper has received funding from UDESC and FAPESC, and from the European Union’s Horizon 2020 research and innovation programme under the LEGaTO Project (legato-project.eu), grant agreement No 780681.
\bibliographystyle{IEEEtran}
\bibliography{draft}

\begin{thebibliography}{10}
\providecommand{\url}[1]{#1}
\csname url@samestyle\endcsname
\providecommand{\newblock}{\relax}
\providecommand{\bibinfo}[2]{#2}
\providecommand{\BIBentrySTDinterwordspacing}{\spaceskip=0pt\relax}
\providecommand{\BIBentryALTinterwordstretchfactor}{4}
\providecommand{\BIBentryALTinterwordspacing}{\spaceskip=\fontdimen2\font plus
\BIBentryALTinterwordstretchfactor\fontdimen3\font minus
  \fontdimen4\font\relax}
\providecommand{\BIBforeignlanguage}[2]{{%
\expandafter\ifx\csname l@#1\endcsname\relax
\typeout{** WARNING: IEEEtran.bst: No hyphenation pattern has been}%
\typeout{** loaded for the language `#1'. Using the pattern for}%
\typeout{** the default language instead.}%
\else
\language=\csname l@#1\endcsname
\fi
#2}}
\providecommand{\BIBdecl}{\relax}
\BIBdecl

\bibitem{Burns:2016}
B.~Burns, B.~Grant, D.~Oppenheimer, E.~Brewer, and J.~Wilkes, ``Borg, omega,
  and kubernetes,'' \emph{Queue}, vol.~14, no.~1, pp. 10:70--10:93, 2016.

\bibitem{Suo-2018}
K.~Suo, Y.~Zhao, W.~Chen, and J.~Rao, ``An analysis and empirical study of
  container networks,'' in \emph{IEEE INFOCOM 2018-IEEE Conf. on Computer
  Communications}.\hskip 1em plus 0.5em minus 0.4em\relax IEEE, 2018, pp.
  189--197.

\bibitem{Rost:2019}
M.~Rost, E.~D\"{o}hne, and S.~Schmid, ``Parametrized complexity of virtual
  network embeddings: Dynamic \& linear programming approximations,''
  \emph{SIGCOMM Comput. Commun. Rev.}, vol.~49, no.~1, pp. 3--10, Feb. 2019.

\bibitem{Souza-2017}
F.~R. de~Souza, C.~C. Miers, A.~Fiorese, M.~D. de~Assun{\c{c}}{\~a}o, and G.~P.
  Koslovski, ``Qvia-sdn: Towards qos-aware virtual infrastructure allocation on
  sdn-based clouds,'' \emph{Journal of Grid Computing}, Mar 2019.

\bibitem{Jupiter-2015}
S.~Arjun, O.~Joon, A.~Amit, A.~Glen, A.~Ashby, B.~Roy \emph{et~al.}, ``Jupiter
  rising: A decade of clos topologies and centralized control in google’s
  datacenter network,'' in \emph{Sigcomm '15}, 2015.

\bibitem{NiranjanMysore:2009}
R.~Niranjan~Mysore, A.~Pamboris, N.~Farrington, N.~Huang, P.~Miri
  \emph{et~al.}, ``Portland: A scalable fault-tolerant layer 2 data center
  network fabric,'' \emph{SIGCOMM Comput. Commun. Rev.}, vol.~39, pp. 39--50,
  2009.

\bibitem{saaty2005making}
T.~L. Saaty, ``Making and validating complex decisions with the {AHP}/{ANP},''
  \emph{J SYST SCI SYST ENG}, vol.~14, no.~1, pp. 1--36, 2005.

\bibitem{Hwang-1981}
C.-L. Hwang and K.~Yoon, \emph{Multiple Attribute Decision Making}.\hskip 1em
  plus 0.5em minus 0.4em\relax Lecture Notes in Economics and Mathematical
  Systems, Springer, 1981.

\bibitem{Guerrero-2018}
C.~Guerrero, I.~Lera, and C.~Juiz, ``Genetic algorithm for multi-objective
  optimization of container allocation in cloud architecture,'' \emph{J GRID
  COMPUT}, vol.~16, no.~1, pp. 113--135, Mar 2018.

\bibitem{guo2018container}
Y.~Guo and W.~Yao, ``A container scheduling strategy based on neighborhood
  division in micro service,'' in \emph{NOMS 2018-2018 IEEE/IFIP Network
  Operations and Management Symp.}\hskip 1em plus 0.5em minus 0.4em\relax IEEE,
  2018, pp. 1--6.

\bibitem{Havet-2017}
A.~Havet, V.~Schiavoni, P.~Felber, M.~Colmant, R.~Rouvoy, and C.~Fetzer,
  ``Genpack: A generational scheduler for cloud data centers,'' in \emph{IEEE
  Int. Conf. on Cloud Engineering (IC2E)}, April 2017, pp. 95--104.

\bibitem{Vaucher-2018}
S.~Vaucher, R.~Pires, P.~Felber, M.~Pasin, V.~Schiavoni, and C.~Fetzer,
  ``{SGX}-aware container orchestration for heterogeneous clusters,'' in
  \emph{IEEE 38th Int. Conf. on Distributed Comp. Systems}, July 2018.

\bibitem{NesiCNSM:2018}
L.~L. {Nesi}, M.~A. {Pillon}, M.~D. {de Assunção}, C.~C. {Miers}, and G.~P.
  {Koslovski}, ``Tackling virtual infrastructure allocation in cloud data
  centers: a gpu-accelerated framework,'' in \emph{2018 14th Int. Conf. on
  Network and Service Management (CNSM)}, Nov 2018, pp. 191--197.

\bibitem{alla2016efficient}
H.~B. Alla, S.~B. Alla, A.~Ezzati, and A.~Touhafi, ``An efficient dynamic
  priority-queue algorithm based on ahp and pso for task scheduling in cloud
  computing,'' in \emph{HIS}.\hskip 1em plus 0.5em minus 0.4em\relax Springer,
  2016, pp. 134--143.

\bibitem{panwartopsis}
N.~Panwar, S.~Negi, M.~M.~S. Rauthan, and K.~S. Vaisla, ``Topsis--pso inspired
  non-preemptive tasks scheduling algorithm in cloud environment,''
  \emph{Cluster Computing}, pp. 1--18.

\end{thebibliography}
\end{document}